\documentclass[%
 english,
 reprint,
 superscriptaddress,
 amsmath,amssymb,
 aps,
 pra,
]{revtex4-1}
\usepackage[T1]{fontenc}
\usepackage[utf8]{inputenc}
\usepackage{amsmath}
\usepackage{graphicx}
\usepackage{babel}
\usepackage{color}

\begin{document}

\title{Anharmonic particles: suppressing van der Waals forces by an external field}

\author{Heino Soo}
\affiliation{4th Institute for Theoretical Physics, Universität Stuttgart, Germany}
\affiliation{Max Planck Institute for Intelligent Systems, 70569 Stuttgart, Germany}
\author{David Dean}
\affiliation{Univ. Bordeaux and CNRS, Laboratoire Ondes et Matière d'Aquitaine (LOMA), UMR 5798, F-33400 Talence, France}
\author{Matthias Krüger}
\affiliation{4th Institute for Theoretical Physics, Universität Stuttgart, Germany}
\affiliation{Max Planck Institute for Intelligent Systems, 70569 Stuttgart, Germany}

\begin{abstract}
We study the classical thermal component of Casimir, or van der Waals, forces between point particles with highly anharmonic dipole Hamiltonians when they are subjected to  an external electric  field. Using a model for which the individual dipole moments saturate in a strong field (a model that mimics the charges in a neutral, perfectly conducting sphere), we find that the resulting Casimir force depends strongly on the strength of the field, as demonstrated by analytical results. For a certain angle between external field and center to center axis, the fluctuation force can be tuned and suppressed to arbitrarily small values. We compare the forces between these anharmonic particles to those between harmonic ones, and also provide a simple formula for asymptotically large external fields, which we expect to be generally valid for the case of saturating dipole moments. 
\end{abstract}
\maketitle

\section{Introduction}

Neutral bodies exhibit attractive forces, called van der Waals or Casimir forces depending on context. The earliest
calculations were formulated by Casimir who studied the force between two metallic parallel plates
\cite{Casimir1948a}, and generalized by Lifshitz \cite{Lifshitz1956}
for the case of dielectric materials.  Casimir and Polder found the force between two polarizable
atoms \cite{Casimir1948}. 
Although van der Waals  forces are only relevant at small (micron scale)
distances, they have been extensively measured, see e.g. \cite{Obrecht2007a,Sukenik1993a}. With recent advances in measurement techniques, including the
MEMS framework \cite{Gad-el-Hak2001}, Casimir-Polder forces become accessible 
 in many other interesting conditions.

Due to the dominance of van der Waals forces in nanoscale devices, there has been much interest in controlling such forces. The full Lifshitz theory for van der Waals forces \cite{Lifshitz1956} shows their dependence on the electrical 
properties of the materials involved.  Consequently the possibility of tuning a material's electric properties opens up the possibility of tuning fluctuation-induced interactions. This principle has been demonstrated in a number of experimental set-ups, for instance  by changing the charge carrier density of materials via laser light \cite{Chen2007,Torricelli2010}, as well as inducing phase transformations by laser heating, which of course engenders a consequent change in electrical properties \cite{Torricelli2010}. There is also experimental
evidence of reduction of van der Waals forces for refractive index-matched colloids \cite{Yethiraj2003,Bonn2009}. 
The question of forces in external fields, electric and magnetic, has been studied in several articles \cite{Robaschik1987,Farina1999,Andrews2006,Salam2006,Wang2006,Tse2012,Cao2013,Dean2016}.
When applying external fields, nonlinear materials (which exhibit ``nonlinear optics'') open up  a variety of possibilities; these possibilities are absent in purely linear systems where the external field and  fluctuating field are merely superimposed. Practically, metamaterials are promising candidates for Casimir force modulation, as they can exhibit strongly nonlinear optical properties
\cite{Kauranen2012,Lapine2014} and their properties can be tuned by external fields \cite{Boardman2011}. The nature and description of fluctuation-induced
effects in nonlinear systems is still under active research  \cite{kheirandish2011finite,Khandekar2014,Makhnovets2016,Soo2016a}. For example, in Ref.~\cite{Soo2016a}, it was shown that nonlinear properties may alter Casimir forces over distances in the nanoscale. However, in the presence of only a small number of explicit examples, more research is needed to understand the possibilities opened up by  nonlinear materials. 

In this manuscript, we consider an analytically solvable model
 for point particles with strongly nonlinear (anharmonic) properties. This is achieved by introducing a maximal, limiting value for the polarization of the particles, i.e., by confining the polarization vector  in anharmonic potential wells. Casimir forces in such systems appear to be largely unexplored, even at the level of two particle interactions. We find that strong external electric fields can be used to completely suppress the Casimir force in such  systems. We discuss the stark difference of forces compared with the case of harmonic dipoles and give an asymptotic formula for the force in strong external fields, which we believe is valid in general if the involved particles have a maximal value for the polarization (saturate).  In order to allow for analytical results, we restrict our analysis to the classical (high temperature) limit. However, similar effects are to  be expected in quantum (low temperature) cases.      
 
We start by computing the Casimir force for harmonic dipoles in an external field in section \ref{sec:Field-dependent-force-between}, where in Sec. \ref{sec:angle}, we discuss the role of the angle between the field and the center to center axis. In section \ref{sec:Model-for-a} we introduce the nonlinear (anharmonic) well model
and compute the Casimir force in an external field in section \ref{sec:Results}. We finally give an asymptotic expression for high fields in section \ref{subsec:Harmonic-approximation-for}.

\section{Force between  harmonic dipoles in a static external field\label{sec:Field-dependent-force-between}}
\subsection{Model}
Classical van der Waals forces can be described by use of quadratic Hamiltonians describing the polarization of the particles involved \cite{boyer1975temperature,Dean2012,Dean2013}. We introduce the system comprising of two dipole carrying particles having the Hamiltonian,
\begin{align}
H^{\left(\mathrm{h}\right)} & =H_{1}^{\left(\mathrm{h}\right)}+H_{2}^{\left(\mathrm{h}\right)}+H_{\mathrm{int}},\label{eq:Hhh}\\
H_{i}^{\left(\mathrm{h}\right)} & =\frac{\mathbf{p}_{i}^{2}}{2\alpha}-\mathbf{p}_{i}\cdot\mathbf{E},\label{eq:H harmonic}\\
H_{\mathrm{int}} & =-2k\left[3\left(\mathbf{p}_{1}\cdot\hat{\mathbf{R}}\right)\left(\mathbf{p}_{2}\cdot\hat{\mathbf{R}}\right)-\mathbf{p}_{1}\cdot\mathbf{p}_{2}\right],\label{eq:Hint}
\end{align}
where $\mathbf{p}_{i}$ is the instantaneous dipole moments of particle $i$. $\alpha$ denotes the 
 polarizability, where for simplicity of presentation, we choose identical particles. The external, homogeneous static electric field $\mathbf{E}$ couples to $\mathbf{p}_{i}$ in the standard manner. The term $H_{\mathrm{int}}$ describes the non-retarded dipole-dipole interaction in $d=3$ dimensions with the coupling constant
\begin{equation}
k=\frac{1}{4\pi\varepsilon_{0}}R^{-3},\label{eq:coupling}
\end{equation}
where $R =|\mathbf{R}|$ with $\mathbf{R}$ the  vector connecting the centers of the two dipoles, while $\hat{\mathbf{R}}$ denotes corresponding unit vector. Since we are considering purely classical forces, retardation is irrelevant. $\varepsilon_{0}$ is the vacuum permittivity, and we use SI units. Inertial terms are irrelevant as well and have been omitted. (Since the interaction does not depend on, e.g., the change of $\mathbf{p}_{i}$ with time, inertial parts can be integrated out from the start in the classical setting.)  

\subsection{Casimir force as a function of the external field}\label{sec:angle}

The force $F$ for the system given in Eqs.~\eqref{eq:Hhh}-\eqref{eq:Hint} can be calculated from (as the external electric field is stationary, the system is throughout in equilibrium)
\begin{align}
F=\frac{1}{\beta}\partial_{R}\ln\mathcal{Z},
\end{align}
where $\mathcal{Z}=\int\mathrm{d}^{3}\mathbf{p}_{1}\int\mathrm{d}^{3}\mathbf{p}_{2}\exp\left(-\beta H\right)$
is the partition function, with the inverse temperature $\beta=1/k_{\mathrm{B}}T$. By using the coupling constant $k$ from Eq. (\ref{eq:coupling}),
this may also be written as 
\begin{equation}
F=\frac{1}{\beta}\left(\partial_{R}k\right)\frac{\partial_{k}\mathcal{Z}}{\mathcal{Z}}.
\end{equation}
Furthermore, we are interested in the large separation limit,
and write the standard series in inverse center-to-center distance (introducing $R\equiv|\mathbf{R}|$) ,
\begin{align}
F & =\frac{1}{\beta}\left(\partial_R k\right)\left(\frac{\partial_{k}\mathcal{Z}}{\mathcal{Z}}\right)_{k=0}\nonumber \\
 & +\frac{1}{\beta}\left(\partial_Rk\right)k\left[\left(\frac{\partial_{k}^{2}\mathcal{Z}}{\mathcal{Z}}\right)-\left(\frac{\partial_{k}\mathcal{Z}}{\mathcal{Z}}\right)^{2}\right]_{k=0}\nonumber \\
 & +\mathcal{O}\left(R^{-10}\right).\label{eq:force_with.dd}
\end{align}
In this series, the first term is of order $R^{-4}$, while the second is of order $R^{-7}$.
The external electric field induces finite (average) dipole moments. For an isolated particle, this is (index $0$ denoting an isolated particle, or $k=0$)
\begin{equation}
\left\langle \mathbf{p}_{i}\right\rangle_0 =\frac{\int\mathrm{d}^{3}\mathbf{p}_{i}\exp\left(-\beta H_{i}\right)\mathbf{p}_{i}}{\int\mathrm{d}^{3}\mathbf{p}_{i}\exp\left(-\beta H_{i}\right)}.\label{eq:average polarization - harmonic}
\end{equation}
For the case of harmonic particles, Eq.~\eqref{eq:H harmonic}, this naturally gives 
\begin{equation}
\left\langle \mathbf{p}_{i}\right\rangle_0=\alpha\mathbf{E}.\label{eq: harmonic <p>}
\end{equation}
The mean dipole moments of the isolated particles in Eq.~(\ref{eq: harmonic <p>}), induced by the external electric field, give rise to a force decaying with $R^{-4}$, i.e., the first term in Eq.~\eqref{eq:force_with.dd}.
This can be  made more explicit by writing
\begin{align}
\left(\frac{\partial_{k}\mathcal{Z}}{\mathcal{Z}}\right)_{k=0} & =2\left\langle \mathbf{p}_{1}\right\rangle _0\cdot\left\langle \mathbf{p}_{2}\right\rangle _0\nonumber \\
 & -6\left(\left\langle \mathbf{p}_{1}\right\rangle _0\cdot\hat{\mathbf{R}}\right)\left(\left\langle \mathbf{p}_{2}\right\rangle _0\cdot\hat{\mathbf{R}}\right).\label{eq:angle}
\end{align}
Representing a  force decaying with $R^{-4}$, this term dominates at large separations. From Eq.~\eqref{eq:angle}, the dependence on the angle between
$\mathbf{E}$ and $\mathbf{R}$ becomes apparent. The induced force can be either attractive
(e.g. $\mathbf{R}\parallel\mathbf{E}$) or repulsive (e.g. $\mathbf{R}\perp\mathbf{E}$) \cite{Jackson1998}. We are aiming at {\it reducing} the Casimir force through the electric field, and thus, term by term, try to obtain small prefactors. The considered term  $\sim R^{-4}$ is readily reduced by choosing $\hat{\mathbf{R}}\cdot\hat{\mathbf{E}}=\frac{1}{\sqrt{3}}$, for which this term is exactly zero, $\left(\frac{\partial_{k}\mathcal{Z}}{\mathcal{Z}}\right)_{k=0}=0$. See the inset of Fig. \ref{fig:force harmonic} for an illustration. In the following sections we will thus study the behavior of the term $\sim R^{-7}$ as a function of the external field, keeping this angle thoughout.

\subsection{Force for the angle $\hat{\mathbf{R}}\cdot\hat{\mathbf{E}}=\frac{1}{\sqrt{3}}$}
For $\hat{\mathbf{R}}\cdot\hat{\mathbf{E}}=\frac{1}{\sqrt{3}}$, the force is of order $R^{-7}$ for large $R$, and reads
\begin{equation}
\left.F\right|_{\hat{\mathbf{R}}\cdot\hat{\mathbf{E}}=\frac{1}{\sqrt{3}}}=\frac{\partial_R k^{2}}{2\beta}\left(\frac{\partial_{k}^{2}\mathcal{Z}}{\mathcal{Z}}\right)_{k=0}+\mathcal{O}\left(R^{-10}\right)\label{eq:aligned force from Z}.
\end{equation}
\begin{figure}
\includegraphics[width=1\columnwidth]{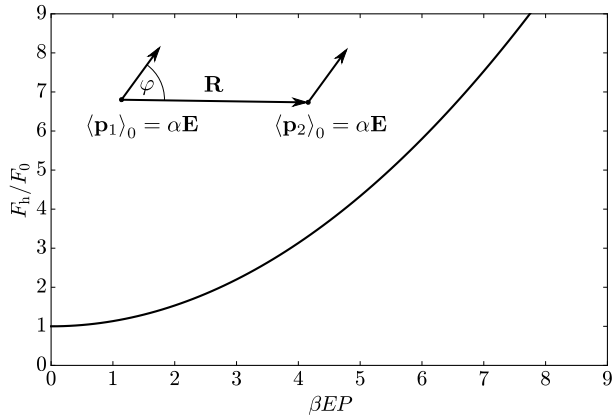} 
\caption{Casimir force between harmonic dipoles as a function of the strength of a static external field. The angle between the field and the center-to-center vector $\mathbf{R}$ is chosen $\varphi=\arccos\left(\frac{1}{\sqrt{3}}\right)$. The ``deterministic'' force $\sim R^{-4}$ then vanishes, so that the force decays as $\sim R^{-7}$.\label{fig:force harmonic}}
\end{figure}
The discussion up to here, including Eq.~\eqref{eq:aligned force from Z}, is valid generally, i.e., for any model describing the individual dipoles. For the case of harmonic dipoles, i.e., for Eq.~\eqref{eq:H harmonic}, we denote $F=F_{\mathrm{h}}$. Calculating $\left(\frac{\partial_{k}^{2}\mathcal{Z}}{\mathcal{Z}}\right)_{k=0}$ for this case yields a result which is partly familiar from the case of harmonic dipoles in the absence of external fields (denoted $F_{0}$), 
\begin{align}
F_{\mathrm{h}}& =\left(1+\frac{2}{3}\alpha\beta E^{2}\right)F_{0}+\mathcal{O}\left(R^{-10}\right),\label{eq:Force.hh}\\
F_{0} & =-\frac{72}{\beta}\left(\frac{\alpha}{4\pi\varepsilon_{0}}\right)^{2}R^{-7}.\label{eq:F_0 harmonic}
\end{align}
Again, for zero field, $E\rightarrow0$, this is in agreement with
the Casimir-Polder force in the classical limit \cite{boyer1975temperature},
given by $F_{0}$. As the field is applied, the force increases, being proportional to
$E^{2}$ for $\alpha\beta E^{2}\gg1$.  This is due to interactions of a dipole induced by the $E$-field with a fluctuating dipole (compare also \eqref{eq:Force_sh} below). The term proportional to $E^2$ is naturally independent of $T$. The force as a function of external field is shown in Fig.~\ref{fig:force harmonic}.

The Casimir force given by Eq. (\ref{eq:Force.hh}) is thus tunable through the external field, but it can only be {\it increased} due to the square power law. While this might be useful for certain applications, we shall in the following investigate the case of highly nonlinear particles. The fact that the force in Eq.~\eqref{eq:F_0 harmonic} is proportional to $\alpha^{2}$ suggests that reduction of the force could be achieved, if the polarizabilities were 
 \emph{dependent on the external field}. In the next section, we will investigate a model for saturating
particle dipole moments, where indeed the forces can be suppressed.
 

\section{Force between saturating dipoles in an external field}

\subsection{Model: Infinite wells \label{sec:Model-for-a}}

\begin{figure}
\begin{centering}
\includegraphics[width=0.8\columnwidth]{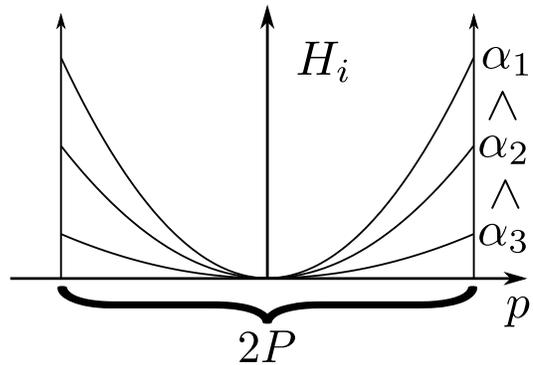} 
\par\end{centering}
\centering{}\caption{Illustration of a simple potential for the individual dipoles, which describes saturation. A parabola of curvature $\alpha^{-1}$ is cut off by a hard ``wall'' at the value $P$. Practically, we simplify even further by letting the polarizability $\alpha$ tend to infinity, so that the potential of Eq.~\eqref{eq:zero-field polarizability} is approached. Physically, $\alpha\to\infty$ means $\alpha\gg\beta P^{2}$. \label{fig:harmonic to well potential}}
\end{figure}

The response of a harmonic dipole to an external field is by construction 
linear for any value of the field (see Eq.~\eqref{eq: harmonic <p>}), and the polarization can be
increased without bound. We aim here to include  saturation by introducing
a limit $P$ for the polarization, such that $\left|\mathbf{p}_{i}\right|<P$ at all times and for all external fields.
This can be achieved by modifying the Hamiltonian in Eq.~(\ref{eq:H harmonic}), asigning an infinite value for $|\mathbf{p}_{i}|>P$. The potential for $|\mathbf{p}_{i}|$ obtained in such a way is illustrated in Fig.~\ref{fig:harmonic to well potential}.   

As we aim to study the effect of saturation, while keeping the number of parameters to a minimum, 
 we additionally
take the limit $\alpha\rightarrow\infty$. This yields an infinite
well potential (see the lower curves of Fig.~\ref{fig:harmonic to well potential} for the approach of this limit),
\begin{equation}
H_{i}^{\left(\mathrm{w}\right)}=\begin{cases}
-\mathbf{p}_{i}\cdot\mathbf{E}, & \left|\mathbf{p}_{i}\right|<P,\\
\infty, & \mathrm{otherwise}.
\end{cases}\label{eq:WD Hamiltonian}
\end{equation}
Such models have been studied extensively in different contexts, as, e.g, asymmetric quantum
wells of various shapes \cite{Rosencher1991,Yildirim2005,Zhang2007},
two-level systems with permanent dipole moments \cite{Bavli1991},
and dipolar fluids \cite{Szalai2009}. These systems are also known
to be tunable with an external electric field \cite{Ahn1987,Tsang1988}.
However the Casimir effect has not been investigated. 

This model for example mimics free electrons confined to a spherical volume, such as in a perfectly conducting, neutral sphere. The charge distribution in a sphere has, additionally to the dipole moment, higher multipole moments, e.g. quadrupolar. For a homogenous external field, the Hamiltonian in Eq.~\eqref{eq:WD Hamiltonian} is however precise, as  higher multipoles couple to spatial derivatives (gradients) of the field \cite{Jackson1998}, and only the dipole moment couples to a homogeneous field. Also, the interaction part, Eq.~\eqref{eq:Hint}, contains in priciple terms with higher multipoles. These do however play no role for the force at the order $R^{-7}$.

\subsection{Polarization and polarizability}
We start by investigating the polarization of an individual particle as a function of the field $E$, resulting from Eq.~\eqref{eq:WD Hamiltonian}, which is defined in Eq.~\eqref{eq:average polarization - harmonic}. It can be found analytically,
\begin{align}
\left\langle \mathbf{p}_{i}\right\rangle _0 & =Q\left(\beta EP\right)P\hat{\mathbf{E}},\label{eq:<p>_E}\\
Q\left(x\right) & =\frac{1}{x}\frac{\left(x^{2}-3x+3\right)e^{2x}-x^{2}-3x-3}{\left(x-1\right)e^{2x}+x+1}.\label{eq:Q(x)}
\end{align}
Note that the product $\beta EP$ is dimensionless. For a small external field, we find the average polarization is given by
\begin{equation}
\left\langle \mathbf{p}_{i}\right\rangle _0  =  \frac{1}{5}\beta P^{2} {\mathbf{E}} + \mathcal{O}\left(E^{3}\right).
\label{eq:zero-field polarizability}
\end{equation}
We hence observe, as expected, that for small field the particles respond harmonically,  with a polarizability $\alpha_0 \equiv \frac{1}{5}\beta P^{2}$. This polarizability depends on temperature, as it measures how strongly the particles' thermal fluctuations in the well are perturbed by the field. We may now give another interpretation of the limit   $\alpha\to 0$ in Fig.~\ref{fig:harmonic to well potential}: in order to behave as a ``perfect'' well, the curvature, given by $\alpha^{-1}$ must be small enough to fulfill  $\alpha\gg\alpha_{0}$.
\begin{figure}
\begin{centering}
\includegraphics[width=1\columnwidth]{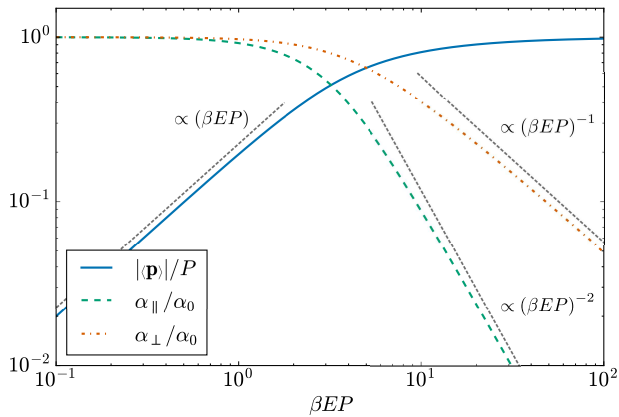} 
\par\end{centering}
\centering{}\caption{Characterization of an isolated particle described by the well model. The mean dipole moment (see Eq. (\ref{eq:<p>_E}))
and polarizations (see Eqs.~(\ref{eq:A_para}) and (\ref{eq:A_perp})). $P$ is the ``width'' of the well potential, and $\alpha_{0}\equiv\frac{1}{5}\beta P^{2}$ denotes  the zero-field polarizability.\label{fig:coefficients.single}}
\end{figure}

The normalized polarization (i.e. $Q\left(\beta EP\right)=\frac{\left|\left\langle \mathbf{p}_{i}\right\rangle _0\right|}{P}$) is shown in Fig.~\ref{fig:coefficients.single} as a function of external field. For small values of $E$, one sees the linear increase, according to Eq.~\eqref{eq:zero-field polarizability}. In the large field limit, the polarization indeed saturates to $P\hat{\mathbf{E}}$. The dimensionless axis yields the relevant scale for $E$, which is given through $(\beta P)^{-1}$. At low temperature (or large $P$), saturation is approached already for low fields, while at high temperature (or low $P$), large fields are necessary for saturation.    

Another important quantity related to the polarization is the polarizability, which is a measure of how
easy it is to induce or change a dipole moment in a system. For harmonic
particles, it is independent of external fields (see Eq. (\ref{eq: harmonic <p>})). 
In the case of anharmonic particles, the field-dependent polarizability tensor $\alpha_{ij}$ is of interest. It is defined through the linear response, 
\begin{equation}
\alpha_{ij}= \frac{\partial \left\langle p_{i}\right\rangle}{\partial E_j}.\label{eq:pol}
\end{equation}
Note that this derivative is  not necessarily taken at zero field $E$, so that $\alpha_{ij}$ is a function of $E$. Indices $i$ and $j$  denote the  components of vectors (in contrast to previous notation). The polarizability tensor as defined in Eq.~\eqref{eq:pol} is measured in the absence of any other particle (in other words, at coupling $k=0$). $\alpha_{ij}$ can be deduced directly from the function $Q$ in Eq.~\eqref{eq:Q(x)}. In general,  we can write
\begin{align}
\alpha_{ij}\left(\beta,E,P\right) & =A_{ij}\left(\beta EP\right)\alpha_{0}.\label{eq:a_para}
\end{align}
Recall the zero-field polarizability $\alpha_{0}\equiv\frac{1}{5}\beta P^{2}$ (Eq. (\ref{eq:zero-field polarizability})).
For the isolated particle, the only special direction is provided by the external field $E$, and it is instructive to examine the polarizability parallel and perpendicular to it. As mentioned, these are related to $Q$, and the corresponding dimensionless amplitudes $A_{\parallel}$ and  $A_{\perp}$ are,
\begin{align}
A_{\parallel}\left(x\right) & = 5\frac{\mathrm{d}}{\mathrm{d}x}Q\left(x\right),
\label{eq:A_para}\\
A_{\perp}\left(x\right) & = 5\frac{1}{x}Q\left(x\right).
\label{eq:A_perp}
\end{align}
The amplitudes for parallel and perpendicular polarizability are also shown in Fig.~\ref{fig:coefficients.single}. The direct connection with the polarization is evident. For small fields, where the polarization grows linearly, the polarizability is independent of $E$. Analytically,
\begin{align}
A_{\parallel}\left(x\right) & =1-\frac{3}{35}x^{2}+\mathcal{O}\left(x^{3}\right),\\
A_{\perp}\left(x\right) & =1-\frac{1}{35}x^{2}+\mathcal{O}\left(x^{3}\right).
\end{align}
For large fields, i.e., when $\beta EP$ is large compared to unity, the polarizability reduces due to saturation effects. Asymptotically for large fields, the polarizability amplitudes
are given as
\begin{align}
A_{\parallel}\left(x\right) & =10x^{-2}+\mathcal{O}\left(x^{-3}\right),\\
A_{\perp}\left(x\right) & =5x^{-1}-10x^{-2}+\mathcal{O}\left(x^{-3}\right).\label{eq:limitA}
\end{align}
The parallel polarizability $\alpha_{\parallel}$ falls off as
$E^{-2}$ and the parallel polarizability $\alpha_{\perp}$ as $E^{-1}$.
The different power laws may be expected, as near saturation, changing the dipoles direction is a softer mode compared to changing the dipoles' absolute value. 
\subsection{Casimir force\label{sec:Results}}

\begin{figure}
\includegraphics[width=1\columnwidth]{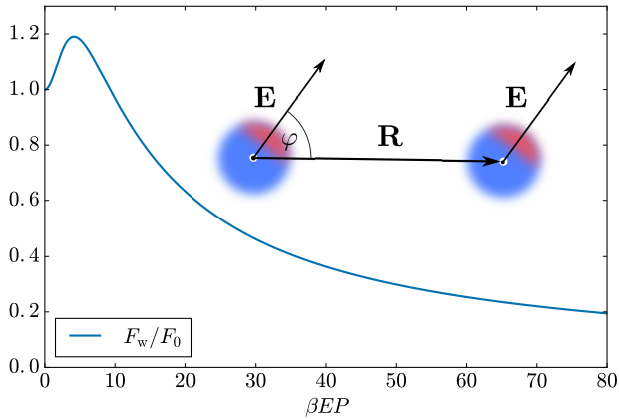} 

\caption{Casimir force between two well particles in an external
electric field $\mathbf{E}$. The angle between the field and the vector $\mathbf{R}$
is $\varphi=\arccos\left(\frac{1}{\sqrt{3}}\right)$.\label{fig:wellplot}}
\end{figure}
The Casimir force between particles described by the well potential, Eq.~(\ref{eq:WD Hamiltonian}), is computed from the following Hamiltonian

\begin{align}
H^{\left(\mathrm{w}\right)} & =H_{1}^{\left(\mathrm{w}\right)}+H_{2}^{\left(\mathrm{w}\right)}+H_{\mathrm{int}},\\
H_{i}^{\left(\mathrm{w}\right)} & =\begin{cases}
-\mathbf{p}_{i}\cdot\mathbf{E}, & \left|\mathbf{p}_{i}\right|<P,\\
\infty, & \mathrm{otherwise},
\end{cases}\label{eq:H harmonic-1}
\end{align}
with the interaction potential $H_{\mathrm{int}}$ given in Eq. \eqref{eq:Hint}. The discussion in section \ref{sec:Field-dependent-force-between} regarding the angle of the external field holds similarly here, i.e., Eq.~\eqref{eq:aligned force from Z} is valid. The force decaying as $R^{-4}$ hence vanishes for the angle $\hat{\mathbf{R}}\cdot\hat{\mathbf{E}}=\frac{1}{\sqrt{3}}$ and we continue by studying the $R^{-7}$ term at this angle. 
Using Eq. (\ref{eq:aligned force from Z}), the 
Casimir force can be found analytically, 
\begin{eqnarray}
F_{\mathrm{w}} & = & f_{\mathrm{w}}\left(\beta EP\right)F_{0}+\mathcal{O}(R^{-10}),\label{eq:Fw}
\end{eqnarray}
with the zero field force
\begin{align}
F_{0}  =  -\frac{72}{\beta}\left(\frac{\alpha_{0}}{4\pi\varepsilon_{0}}\right)^{2}R^{-7},
\end{align}
and the dimensionless amplitude
\begin{align}
f_{\mathrm{w}}\left(x\right)  = \frac{25}{3}\frac{1}{x^{4}}\frac{\left(x^{2}+3\right)\sinh\left(x\right)-3x\cosh\left(x\right)}{\left[x\cosh\left(x\right)-\sinh\left(x\right)\right]^{2}}\\
  \times\left[\left(2x^{2}+21\right)x\cosh\left(x\right)-\left(9x^{2}+21\right)\sinh\left(x\right)\right].\nonumber 
\end{align}
Again,  $\alpha_{0}\equiv\frac{1}{5}\beta P^{2}$ is  the zero-field
polarizability (see Eq. (\ref{eq:zero-field polarizability})).
The force is most naturally expressed in terms of $F_{0}$, which is the force at zero-field, equivalent to Eq.~(\ref{eq:F_0 harmonic}). The amplitude $f_\mathrm{w}$ is then dimensionless and depends, as the polarization, on the dimensionless combination $\beta EP$.

The force is shown in Fig. \ref{fig:wellplot}. For zero external fields, the curve starts at unity by construction, where the force is given by $F_0$. The force initially increases for small values of $\beta EP$, in accordance with our earlier analysis of harmonic dipoles. After this initial regime of linear response,  the Casimir force {\it decreases} for $\beta EP\agt 1$, and, for $\beta EP\gg 1$, asymptotically approaches zero with $E^{-1}$,  
\begin{align}
F_{\mathrm{w}} & =-\frac{48P^{3}}{\left(4\pi\varepsilon_{0}\right)^{2}}R^{-7}E^{-1}+\mathcal{O}\left(E^{-2}\right).\label{eq:limit}
\end{align}
This behavior yields enormous potential for applications: by changing the external field, the force can be switched on or off. 

\begin{figure}
\includegraphics[width=1\columnwidth]{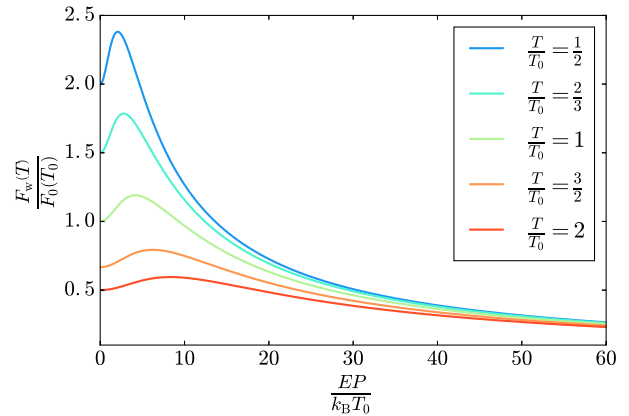} 
\caption{Temperature dependence of the Casimir force for well particles. For small $E$, the force decreases with temperature, because the zero-field polarizability is $\alpha_{0}=\frac{1}{5}\beta P^{2}$. For large $E$, the force is unexpectedly independent of $T$. \label{fig:well force temperature}}
\end{figure}

The asymptotic law in Eq.~\eqref{eq:limit} gives another intriguing insight: for large field, the force is {\it independent} of temperature. This is in contrast to the fact that (classical) fluctuation induced forces in general do depend on temperature. This peculiar observation is a consequence of cancellations between factors of $\beta$, and might yield further possibilities for applications. This is demonstrated in Fig.~\ref{fig:well force temperature}. Indeed, we see that for small values of $E$, the force does depend on temperature, while for large fields, the  curves for different values of temperature fall on top of each other. As a remark, we note that $F_0$ is inversely proportional  to temperature, in contrast to $F_0$ for harmonic particles in Eq.~\eqref{eq:F_0 harmonic}. This is because the zero field polarizability depends on temperature for the well potentials considered here.  


\subsection{Asymptotic formula for high fields\label{subsec:Harmonic-approximation-for}}

What is the physical reason for the decay of the force for large field $E$ observed in Fig.~\ref{fig:wellplot}? For large values of $\beta EP$, the force may be seen as an interaction between a stationary dipole and a fluctuating one. 
This is corroborated by a direct computation of the force between a stationary dipole $\mathbf{q}$, pointing in the direction of the electric field, and a particle with Hamiltonian
\begin{align}
H_{1}^{\left(\mathrm{s}\right)} & =\frac{p_{\parallel}^{2}}{2\alpha_{\parallel}}+\frac{\mathbf{p}_{\perp}^{2}}{2\alpha_{\perp}}-\mathbf{p}\cdot\mathbf{E},
\end{align}
where ``perpendicular'' and ``parallel'' refer to the direction of the $E$ field as before. The two such hypothetical  particles interact via the Hamiltonian  
\begin{align}
H_{\mathrm{int}}^{\left(\mathrm{s}\right)} & =-2k\left[3\left(\mathbf{p}\cdot\hat{\mathbf{R}}\right)\left(\mathbf{q}\cdot\hat{\mathbf{R}}\right)-\mathbf{p}\cdot\mathbf{q}\right].
\end{align}




Choosing the angle between $\bf R$ and $\bf E$ as before, we find for the force between these particles (to leading order in $k$), 
\begin{align}
F_{\mathrm{s}} & =-24\alpha_{\perp}q^{2}\left(\frac{1}{4\pi\varepsilon_{0}}\right)^{2}R^{-7}.\label{eq:Force_sh}
\end{align}
This result can be related to Eq.~\eqref{eq:limit}. Substituting  $\mathbf{q}=P\hat{\mathbf{E}}$, the value at saturation, and $\alpha_{\perp}=5/(\beta EP)\alpha_{0}=P/E$ (using the leading term for large field from Eq.~\eqref{eq:limitA}), we find 
\begin{align}
F_{\mathrm{s}} & =-24\frac{P^3}{E}\left(\frac{1}{4\pi\varepsilon_{0}}\right)^{2}R^{-7}.\label{eq:Force_sh_2}
\end{align}
This is identical to Eq.~\eqref{eq:limit}, except for a factor of two. This is expected, as this factor of two  takes into account the force from the first fixed dipole interacting with the second fluctuating
one and vice versa. We have thus demonstrated that Eq.~\eqref{eq:Force_sh} may be used to describe the behavior of the force for large values of $E$. The importance of this observation lies in the statement, that such reasoning might be applicable more generally: in case of more complex behavior of $p(E)$, i.e., more complex (or realistic) particles. We believe that the value of $q$ at saturation and the polarizability $\alpha_\perp$ near saturation can be used to accurately predict the force in the limit of large external fields via Eq.~\eqref{eq:Force_sh}.      



\section{Summary}

We have demonstrated how the classical Casimir-Polder
force between two saturating dipoles can be suppressed by applying  an external
static electric field. Of special interest is the angle $\varphi=\arccos\left(\frac{1}{\sqrt{3}}\right)$
between the external field and the vector connecting the dipoles,
for which the deterministic dipole-dipole interaction vanishes. The remaining ``Casimir-Polder'' part can then be tuned and is arbitrarily suppressed at large values of external fields due to the vanishing polarizability. The force in that case decays with $E^{-1}$. This is in strong
contrast to harmonic dipoles, which experience
an increase of the force in the presence of an external field, growing with $E^{2}$. We also provided a simple formula to estimate the force between particles under strong fields. It would be interesting to extend the results here to macroscopic objects composed of such dipole carrying particles, here multi-body effects will potentially change the physics for dense systems. However for dilute systems, where the pairwise approximation of van der Waals forces is accurate, the results 
obtained here are directly applicable and thus the modulation of Casimir or van der Waals forces, predicted here 
will apply to a certain extent. Of course an important main difference in more than two body systems is that the deterministic  component of the interaction cannot be obviously cancelled by a uniform electric field, as there is more than one center to center vector, denoted by $\mathbf{R}$ in this paper, separating the interacting dipoles.  

We thank G. Bimonte, T. Emig, N. Graham, R. L. Jaffe and M. Kardar for useful discussions during early stages of this work.
This work was supported by Deutsche Forschungsgemeinschaft (DFG) grant No. KR 3844/2-1 and MIT-Germany Seed Fund Grant No.~2746830.

%

\end{document}